\documentclass{svjour3}                     
\smartqed  
\usepackage{graphicx}
\usepackage{mathptmx}      
\usepackage{natbib}

\journalname{SSRv}
\begin{document}

\title{Neutral H density at the termination shock: a consolidation of recent results}

\titlerunning{Neutral H density at TS}        

\author{M. Bzowski         \and
        E. M{\"o}bius      \and
        S. Tarnopolski     \and
        V. Izmodenov       \and
        G. Gloeckler
}

\authorrunning{Bzowski et al.} 

\institute{M. Bzowski \at
              Space Research Centre PAS, Bartycka 18A, Warsaw, Poland \\
              \email{bzowski@cbk.waw.pl}           
           \and
           E. M{\"o}bius \at
              Space Science Center and Department of Physics, University of New Hampshire, Durham NH
           \and
           S. Tarnopolski \at
              Space Research Centre PAS, Warsaw, Poland
           \and
           V. Izmodenov \at
              Moscow State University and Space Research Institute RAS, Moscow, Russia
           \and 
           G. Gloeckler \at
               Department of Atmospheric, Oceanic, and Space Science, University of Michigan, Ann Arbor MI
}

\date{Received: date / Accepted: date}

\maketitle

\begin{abstract}
 We discuss a consolidation of determinations of the density of neutral interstellar H at the nose of the termination shock carried out with the use of various data sets, techniques, and modeling approaches. In particular, we focus on the determination of this density based on observations of H pickup ions on Ulysses during its aphelion passage through the ecliptic plane. We discuss in greater detail a novel method of determination of the density from these measurements and review the results from its application to actual data. The H density at TS derived from this analysis is equal to $0.087 \pm 0.022$~cm$^{-3}$, and when all relevant determinations are taken into account, the consolidated density is obtained at $0.09 \pm 0.022$~cm$^{-3}$. The density of H in CHISM based on literature values of filtration factor is then calculated at $0.16 \pm 0.04$~cm$^{-3}$.
\end{abstract}

\section{Introduction}
\label{sec:intro}
With the plethora of detailed measurements available from heliospheric missions such as Voyager, Ulysses, SOHO, Cassini, ACE, Wind, and many others and in anticipation of the first mission dedicated to remote-sensing studies of the heliospheric interface IBEX, a consolidation of heliospheric parameters is an important task. Key parameters in this quest are the density of neutral interstellar hydrogen and helium at the nose of the termination shock. While a consensus was reached on the parameters of interstellar helium already a few years ago \citep{mobius_etal:04a, mobius_etal:05a, witte:04, gloeckler_etal:04a, vallerga_etal:04a, lallement_etal:04a}, the hydrogen density remained more uncertain until very recently due to more complex effects on this species within the heliosphere. A dedicated team supported by the International Space Science Institute from Bern, Switzerland, has recently reached a consolidation of the density of neutral interstellar hydrogen at the nose of the termination shock based on different data sets and modeling. We discuss this consolidation, focusing mostly on the determination of the density at the termination shock based on hydrogen pickup ion (H PUI) observations by Ulysses during its passage through the ecliptic plane at aphelion of its orbit, performed using a novel analysis method.
\section{Neutral gas density from pickup ion observations}
\label{sec:hydroDensi}
\subsection{Past attempts to derive the density at the termination shock}
\label{sec:pastAttempts}
Pickup ions are neutral interstellar atoms which become ionized and picked up by the frozen-in magnetic field of the solar wind. A simple model of the transport of pickup ions in the solar wind was developed by \citet{vasyliunas_siscoe:76} under assumption that (i) the atoms of the seed (neutral) population are stationary with respect to the Sun at the moment of ionization; (2) the solar wind expands radially away from the Sun with a velocity $V_{\mathrm{SW}}$, is spherically symmetric and stationary, and the frozen-in magnetic field is perpendicular to the expansion direction -- thus there is no net parallel streaming of the PUI population (the PUI fluid expands radially); (3) the population of newly-injected pickup ions undergoes an immediate isotropization of the pitch angle with respect to the local magnetic field direction, forming a spherical shell distribution; (4) the evolution of the pickup ion distribution function with distance is adiabatic; and (5) the ionization rate of the seed population is spherically symmetric and invariable with time. 

With these assumptions, the distribution function of pickup ions at a heliocentric distance $r$ at the upwind axis is described by the formula:
\begin{equation}
\label{equ:01}
f_{pui}\left(r,w\right) = \frac{3}{8 \pi V_{\mathrm{SW}}^4} \beta_{\mathrm{ion}}\left(\frac{r_E}{r}\right)^2 \left(r w^{-3/2}\right) n_{\mathrm H}\left( r w^{-3/2}\right),
\end{equation}
where $w = v/V_{\mathrm{SW}}$, $v$ is the specific velocity of a pickup ion in the co-moving frame of the solar wind, and
\begin{equation}
\label{equ:02}
\beta_{\mathrm{ion}} = \beta_{\mathrm{ion,E}} \left(r/r_E\right)^2,
\end{equation}
where $\beta_{\mathrm{ion,E}}$ is the production rate of pickup ions at the reference distance $r_E$, usually adopted as equal to 1~AU. The production rate is assumed to be equal to the loss rate of the interstellar gas, which implies that $n_{\mathrm H}$ is also a function of $\beta_{\mathrm{ion}}$. 

It is evident from Eq.(\ref{equ:01}) that the distribution function observed at a location $\left(r, \theta \right)$ for a given velocity $v$  corresponds to the density of the seed population at a distance from the Sun equal to $r w^{-3/2}$. Thus, the absolute spectrum of pickup ions in velocity at a given location in space yields the density profile of the interstellar gas seed population with distance from the Sun if the PUI production rate and the speed of the solar wind are known. With the local density of the neutral gas in hand, one can extrapolate this quantity to the termination shock providing a sufficiently accurate model of the depletion of neutral interstellar gas in the inner heliosphere is available. Such a method has been used to infer the density of interstellar H at the termination shock by \citet{gloeckler:96a} and \citet{gloeckler_geiss:01a}. A similar method was used by \citet{mobius_etal:88a} to determine the termination shock density of interstellar He from observations at 1~AU.

In some cases, an alternative method can be used. Since He$^{+}$ pickup ions are solely created by charge exchange between the neutral He atoms and solar wind alphas, and since Ulysses SWICS is able to simultaneously observe the He$^{2+}$ PUI and solar wind fluxes at $\sim ~5$~AU, where neutral interstellar He is not significantly depleted, \citet{gloeckler_etal:97} could obtain the local He density  with an uncertainty that depends only on the accuracy of the knowledge of charge exchange cross section and does not depend on the absolute calibration of the instrument. This method, however, is not directly applicable to H because the production of pickup ions depends on two comparable ionization channels, i.e. charge exchange with solar wind and EUV ionization. Consequently, the rate cannot be easily eliminated from the analysis by simultaneous measurements. The remaining uncertainty in the total ionization rate translates into the uncertainty in the density and its significance is illustrated by the fact that \citet{gloeckler_geiss:01a} had to invoke a high electron ionization rate of $2.4\times 10^{-7}$~s$^{-1}$ at 1~AU to explain the pickup ion velocity distribution, while current estimates for the time interval of the observations suggest that it is equal to about $0.8\times 10^{-7}$~s$^{-1}$ \citep{bzowski:08a}.

\subsection{Modeling uncertainties in pickup ion interpretations}
\label{sec:modelUncert}

Because the aforementioned method to deduce the density at the termination shock depends on the knowledge of the neutral gas density inside the observer location and interplanetary parameters greatly influence the relation between the local density and the density at the shock, it is rather sensitive to modeling choices and the knowledge of parameters, such as ionization rates and radiation pressure. The transport of PUI from their birthplace to the detector is quite complex \citep[see, e.g.][]{isenberg:87, chalov_fahr:98b, chalov:06, malama_etal:06} and the attenuation processes of interstellar gas, including ionization and radiation pressure, are strongly anisotropic and time-dependent on time scales from days to decades \citep[see, e.g.,][]{bzowski:08a}. Thus, for an instantaneous distribution of PUI (1) one cannot assume the local production rate of PUI is equal to the loss rate of the interstellar gas, which is one of the factors shaping $n_{\rm H}\left(r, \theta\right)$, (2) one cannot be sure where exactly the boundary of the cavity is located because it changes with time, and (3) one can use neither the cold model, nor even the classical hot model \citep{fahr:78, fahr:79, thomas:78, wu_judge:79a, lallement_etal:85b} for the local density of interstellar gas. The reasons for this, partly discussed already by \citet{rucinski_bzowski:96}, are as follows. Firstly, H is subject to a complex behavior of the radiation pressure, which depends on the radial velocity of an atom \citep{tarnopolski_bzowski:08b, bzowski:08a} and is proportional to the net solar flux in the Lyman-$\alpha$ line. Its magnitude varies with the phase of the solar cycle and over a solar rotation \citep[e.g.,][]{bzowski:01a, bzowski:08a}. Secondly, the ionization rate is composed of charge exchange with the solar wind protons, photoionization \citep{bzowski:01a, auchere_etal:05a, bzowski_etal:08a, bzowski:08a}, and electron-impact ionization \citep{bzowski_etal:08a, bzowski:08a}. All three contributors vary both with time \citep{bzowski:01a, bzowski:08a} and with heliolatitude \citep{mccomas_etal:99, bzowski_etal:03, quemerais_etal:06b}. 
\citet{gloeckler_geiss:01a} have determined the total ionization rate that includes all of these components from the slope of the observed PUI distribution. However, this method works under the stated assumption that at least over the observation period the PUI production rate is equal to the loss rate of the interstellar neutrals. Also it should be noted, the modeling of the PUI distributions for all comparisons with observations, except if just the PUI flux at the observer location, i.e. at the PUI cut-off is used, is tied to the reasonable assumption that the full adiabatic cooling in the expanding solar wind \citep{vasyliunas_siscoe:76} is valid, and assumption that has not been independently tested with observations.  
Finally, there is a strong departure of the hydrogen distribution function at the termination shock from a shifted-Maxwellian due to the filtering processes in the heliospheric interface region \citep{baranov_etal:98a, izmodenov:01, izmodenov_etal:05a, heerikhuisen_etal:06a, izmodenov_baranov:06a}. 

\begin{figure}
  \includegraphics{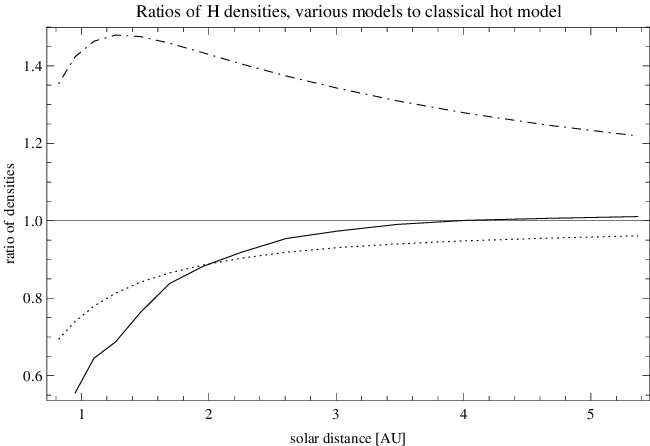}
\caption{Ratios of densities of neutral interstellar H computed using various models relative to the results of the classical hot model. Shown is the profile along the Ulysses-Sun line in the solar equatorial plane towards the Ulysses aphelion. Solid line: Moscow MC model (i.e., conditions at the termination shock non-Maxwellian and non-homogeneous); dotted line: radiation pressure sensitive to the radial velocity of the atoms (non-flat solar Lyman-$\alpha$ line shape adopted); dash-dot: full time-dependent treatment of the radiation pressure and the ionization rate.}
\label{fig:1}       
\end{figure}

An illustration how strongly including or excluding different physical processes in the modeling can influence the simulated density profile is shown in Fig. \ref{fig:1}. It shows simulated densities as a function of distance from the Sun in the equatorial plane towards the aphelion position of Ulysses obtained with three different modeling approaches normalized to the density profile from the classical hot model of the interstellar gas.

As can be seen in Fig. \ref{fig:1}, at heliospheric locations relevant to the Ulysses in-ecliptic measurements, various modeling strategies lead to differences in the predicted gas density, reaching a 30\% level at $\sim 5$~AU and increasing towards the Sun. The magnitude of variations shown here should be regarded as a lower boundary because the model used does not take into account the variations in the distribution function in the source region, i.e. at the termination shock, which were shown to amount to about 10\% by \citet{izmodenov_etal:05b} and \citet{izmodenov_etal:08a}. Therefore, the classical hot model can only be regarded as a qualitative representation of the true density, and the quality of the approximation is reduced with a decreasing distance from the Sun. The model distribution function derived from the classical hot model is also affected by the inherent deficiencies of this model, of course, and its direct use to the interpretation of PUI measurements is not recommended. 

\begin{figure}
  \includegraphics{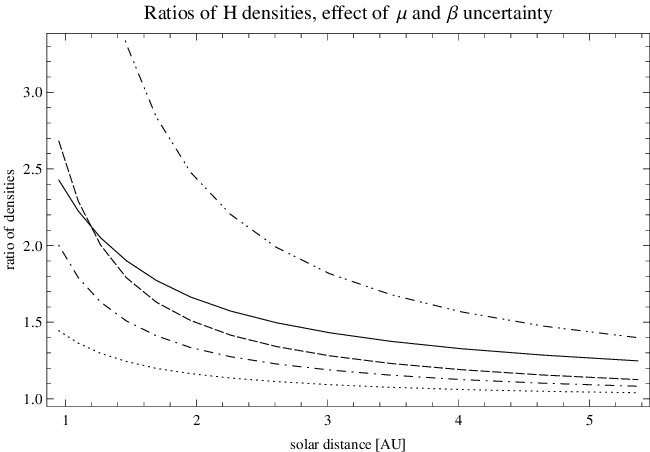}
\caption{Effects of changes in the absolute calibration of the ionization rate and solar Lyman-$\alpha$ flux on the neutral gas density profile. Shown are ratios of densities computed using the hot model for reduced $\mu$ and $\beta$ values over nominal values obtained from recent measurements. Solid line: $\beta$ reduced by 25\%; dots, dash-dot, dashes: $\mu$ reduced by 10\%, 20\%, and 30\%, respectively; dash-dot-dot: both $\mu$ and $\beta$ reduced, respectively, by 25\% and 30\%. }
\label{fig:2}
\end{figure}

Another factor relevant for the calculation of the absolute value of the H PUI distribution function is the accuracy of the heliospheric parameters such as the solar wind flux, solar EUV flux, solar Lyman-$\alpha$ flux, and the shape of the solar Lyman-alpha line. The absolute calibration of the solar wind flux measured by various instruments has been known with an accuracy of about 25\% (cf the OMNI-2 data base, \citet{king_papitashvili:05}, and an early review by \citet{bzowski:01b}). Similarly, the absolute calibration of the flux in the solar Lyman-alpha line has changed within the past 30 years by about 30\% \citep[see][]{ vidal-madjar:75, woods_rottman:97, tobiska_etal:00c, woods_etal:00, floyd_etal:02a} and even now is known to no better than 10\%. The influence of uncertainties in these parameters on the absolute densities of neutral interstellar hydrogen inside the Ulysses orbit near ecliptic is shown in Fig. \ref{fig:2}. 

As evident from Fig. \ref{fig:2}, uncertainties in the absolute ionization rate of 25\% translate into uncertainties of $\sim 25$\% in the density of interstellar gas at Ulysses at $\sim 5$~AU, and an uncertainty of 30\% in radiation pressure leads to a density uncertainty of 10\% at 5~AU. Both uncertainties increase dramatically towards the Sun. All these uncertainties have to be included in the determination of the H density at the termination shock from PUIs in the inner heliosphere.

\subsection{Robust method to determine the density at the termination shock }
\label{sec:robust}

Recently, \citet{bzowski_etal:08a} have presented a novel method to determine the density of interstellar H at the nose of the termination shock using measurements performed at the boundary of the cavity of neutral interstellar gas in the inner heliosphere. 

\subsubsection{Simplified analytic derivation}
\label{sec:simpliAna}
Based on the distribution function of H PUI as proposed by \citet{vasyliunas_siscoe:76} and repeated in Eq. (\ref{equ:01}), the phase space density at $w = 1$ (i.e., for locally injected ions) is directly proportional to the local source function of pickup ions $S\left(r\right)$, defined as:
\begin{equation}
\label{equ:03}
S\left(r\right) = \beta_{\mathrm{ion}}\,n_{\mathrm H}\left(r, \beta_{\mathrm{ion}}\left(r\right)\right).
\end{equation}
Using a cold model of neutral interstellar hydrogen \citep{fahr:68, axford:72} and assuming perfect compensation of the solar radiation pressure and gravity, the gas density $n\left(r\right)$ on the upwind axis varies as:
\begin{equation}
\label{equ:04}
n\left(r\right) = n_0\, \exp\left[-\beta_{\mathrm{ion}} \frac{r}{V_{\mathrm{ISM}}} \right]
\end{equation}
The local PUI production rate is equal to:
\begin{equation}
\label{equ:05}
S\left( r \right) = \beta_{\mathrm{ion}}\, n_0\, \exp\left[-\frac{\beta_{\mathrm{ion}}\, r_E^2}{V_{\mathrm{ISM}}\,r} \right] 
\end{equation}
and scales linearly with the density at the termination shock. Now we seek a distance $r_0$ such that the source function does not depend on $\beta_{\mathrm{ion}}$, requiring that: 
\begin{equation}
\label{equ:06}
\frac{\partial S\left(r_0\right)}{\partial \beta_{\mathrm{ion}}} = 0 = n\left(r_0, \beta_{\mathrm{ion}}\left( r_0 \right)\right) + \beta_{\mathrm{ion}}\left( r_0\right)\, \frac{\partial n \left(r_0, \beta_{\mathrm{ion}}\right)}{\partial \beta_{\mathrm{ion}}}.
\end{equation}
Combining Eq.(\ref{equ:04}) and Eq.(\ref{equ:06}) we get:
\begin{equation}
\label{equ:07}
n\left(r_0\right) = -\beta_{\mathrm{ion}}\left(r_0\right)\,  \frac{\partial n \left(r_0, \beta_{\mathrm{ion}}\right)}{\partial \beta_{\mathrm{ion}}}.
\end{equation}
With
\begin{equation}
\label{equ:08}
\frac{\partial n}{\partial \beta_{\mathrm{ion}}} = -n\, \frac{r}{V_{\mathrm{ISM}}},
\end{equation}
the distance $r_0$ becomes:
\begin{equation}
\label{equ:09}
r_0 = \frac{V_{\rm ISM}}{\beta_{\mathrm{ion}}}\,\,\, {\rm{ or }}\,\,\, r_0 = \frac{\beta_E\, r_E^2}{V_{\mathrm{ISM}}}
\end{equation}
in connection with Eq. (\ref{equ:02}). Inserting this result to Eq. (\ref{equ:04}), we find that the exact location where the source function is insensitive to the ionization rate is found precisely at the boundary of the heliospheric cavity, i.e. where the density is decreased to 1/e of the value at the termination shock. To calculate $n_0$ from the locally observed PUI source strength $S\left(r_0\right)$ the ionization rate must be determined. However, as long as the observations are made close to $r_0$ moderate uncertainties in the knowledge of the ionization rate only result in rather small uncertainties in the determination of the termination shock density $n_0$. In a comparison between different simulations and Ulysses PUI data \citet{bzowski_etal:08a} showed that this method is also robust against uncertainties in the radiation pressure and against variations in the modeling approaches.

\subsubsection{Simulations of the H PUI production rate at Ulysses}
\label{sec:simulUly}
\citet{bzowski_etal:08a} took a subset of the Ulysses H PUI data obtained over 13 months in 1997 and 1998 during the Ulysses passage through the ecliptic plane at aphelion and averaged the observed distribution function to eliminate fluctuations of the local heliospheric conditions. The resulting local production rate of H PUI at Ulysses was equal to $7.26\times10^{-10}$~cm$^{-3}$~s$^{-3}$. 

The simulations  to caclulate the density of neutral hydrogen in the inner heliosphere were performed with the use of a combination of two models. First, the density of interstellar hydrogen at the nose of the termination shock was calculated using the Moscow MC model  \cite{baranov_malama:93} with solar wind alphas and interstellar He$^+$ ions included \citep{izmodenov_etal:03b}. This density was split into two populations, primary and secondary, and their respective bulk velocities and temperatures were registered. The parameters of the two populations were subsequently handed over to the Warsaw kinetic time- and latitude-dependent model \citep{rucinski_bzowski:95b, bzowski_etal:97, bzowski_etal:02, tarnopolski_bzowski:08b}, which calculated a time series of densities and PUI production rates from the primary and secondary populations at 16 evenly-spaced moments during the observation interval at distances and offset angles coinciding with the Ulysses positions. The production rates were averaged over the appropriate time periods so that they could be compared with the value obtained from observations. 

\citet{bzowski_etal:08a} started with a test whether the handover of the calculations from the Moscow MC to the Warsaw kinetic model is robust. The point was that the distribution function of the neutral gas at the termination shock is not independent of the offset angle from the upwind direction, but the Warsaw model makes this assumption. This check was made by running both models with identical parameters, assuming that radiation pressure and ionization rate are averaged over the solar cycle, spherically symmetric, and invariable with time. The version of the MC model used in the simulations is stationary and axially symmetric. The Warsaw model was de-rated to similar assumptions and run with the parameters of the heliospheric gas populations as predicted by the Moscow model at the nose of the termination shock. Finally, the densities predicted by both models along a radial line connecting Ulysses with the Sun were compared. It was discovered that for this geometry, discrepancies between the results begin at $\sim 3$~AU from the Sun and increase inward, but at the Ulysses locations, the two models agree within $\sim 1$\%. Hence, using the static and axially symmetric MC model to determine the parameters of the gas at the termination shock and handing these parameters to a time-dependent, full 3D Warsaw kinetic model was a reasonable and unbiased approach.

\citet{bzowski_etal:08a} studied in detail variations of the solar wind density and velocity as a function of time and heliolatitude as well as radiation pressure, photoionization and electron-impact rates for the observations time. They developed a time- and heliolatitude-dependent model of the charge exchange ionization rate and a spherically symmetric, time-dependent models of the photoionization and electron-ionization rates, as well as a time- and radial velocity-dependent model of radiation pressure, covering a large time span straddling the interval of Ulysses measurements, and implemented them all in the Warsaw code. With those in hand, they calculated the PUI production rates at Ulysses for the 16 observation periods and averaged the result. 

The robustness of this result was verified in a multitude of ways. First, tests of the robustness against modeling strategies were performed. The reaction of the simulated PUI production rate to a down-grading of the model from its full 3D time dependence was tested. First, the fully time-dependent ionization rate and radiation pressure were replaced with static values relevant for the 16 observation points. Then, the sensitivity of the radiation pressure to radial velocities of the atoms was switched off. Tests of the latitude dependence in the ionization rate were performed.  

\begin{figure}
  \includegraphics{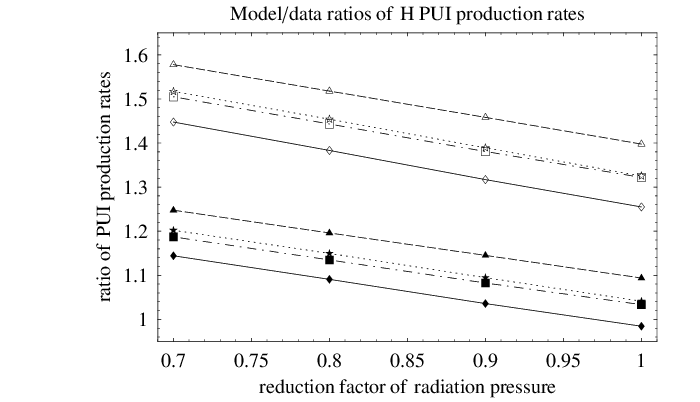}
\caption{Model-to-data ratios of the source function of the H pickup ions at Ulysses as a function of  reduction factor of the net solar Lyman-alpha flux used in the simulation. The group of lines with filled symbols corresponds to the simulations performed with the density at the nose of the termination shock equal to 0.095~cm$^{-3}$, the open symbols correspond to 0.117~cm$^{-3}$. Diamonds correspond to the nominal values of the ionization rate and radiation pressure dependent on $v_{r}$, stars to the values of the ionization rate reduced by 25\% and radiation pressure dependent on $v_{r}$, and squares and triangles to radiation pressure independent on $v_{r}$ with the ionization rate, respectively, on the nominal level and reduced by 25\%. }
\label{fig:3}       
\end{figure}

With this result in hand, the gas density at the termination shock was computed from the relation:
\begin{equation}
\label{equ:13}
n_{\mathrm{TS}} = n_{\mathrm{TS,sim}} \, \frac{S_{\mathrm{obs}}}{S_{\mathrm{sim}}}
\end{equation}
where the index ``sim'' refers to a simulated quantity and ``obs'' to an observed value. The magnitudes of the combined effects are shoewn in Fig. \ref{fig:3} \citep[from][]{bzowski_etal:08a}, which presents the combined variation of all modeling parameters in comparison with the observations. It is evident that the calculated PUI production rate at Ulysses only weakly depends on these variations.

Each of the simulations was repeated with departures of the ionization rate and radiation pressure from their nominal (measured) values. In particular, the radiation pressure was reduced by a factor of 0.9, 0.8, and 0.7 (i.e., by 10\%, 20\%, and 30\%), and the ionization rate by 25\%. The local PUI production rate changed just by a few percent in the case of the ionization rate variation. The effect of radiation pressure reduction was more pronounced, but still within $\sim 15$\%. 

Finally, it was checked whether the conclusions hold when the parameters in the LIC are changed. The entire simulation program was repeated with a different set of parameters, which introduced the simulated density at the termination shock increased by $\sim 25$\%; the most important difference, however, was a totally different breakdown between the primary and secondary populations. The simulated production rates increased almost exactly by the same percentage (as can be seen from the fact that the pairs of lines between the two groups shown in Fig. \ref{fig:3} are almost perfectly parallel), proving that indeed, the PUI source function at Ulysses scales linearly with the density at the termination shock.

In a final test, the simulations were repeated with a different coupling factor between the protons and neutral atoms. It could be shown that replacing the charge exchange cross-section formula from \citet{maher_tinsley:77} with that by  \citet{lindsay_stebbings:05a} does not visibly affect the results.

\section{Determination of hydrogen density at the termination shock}
\label{sec:determ}

To determine the density of neutral H at the nose of the termination shock, \citet{bzowski_etal:08a} took the simulated values of the production rate at Ulysses relevant for the current calibration of the ionization rate and radiation pressure with their uncertainty bars (which correspond to the points for the reduction factor equal to 0.9 or 1.0 in Fig. \ref{fig:3}) and calculated $n_{\mathrm{TS}}$ from the observed PUI fluxes at the Ulysses aphelion. Since the modeling uncertainty was just a few percent, almost the entire error in the result is due to the uncertainty of determination of the PUI production rates from Ulysses observations (mostly the geometric factor of the instrument), which is known to about 20\%. This analysis  brought the density of interstellar H at the termination shock to 0.087~cm$^{-3} \pm 25$\%.

\section{Consolidation of the interstellar H density}
\label{sec:consol}
\subsection{Density of H at the termination shock}

Determinations of the neutral H density at the termination shock, obtained using several different observation methods, were recently compared and consolidated in the framework of an International Space Science Institute Team. These values were then used to infer the H density in the CHISM based on a global heliosphere modeling.

As discussed in the previous section, a re-analysis of the Ulysses SWICS H PUI observations resulted in a H density at the termination shock of $n_{\mathrm{H,TS}} = 0.087 \pm 0.022$~cm$^{-3}$, where all observational, modeling, and external parameter uncertainties are included in the overall uncertainty of 25\% \citet{bzowski_etal:08a}. The neutral H density at the TS was also determined using two independent methods and data sources.

One of them was an analysis of the solar wind slowdown resulting from its mass-loading due to the accumulation of H pickup ions. \citet{richardson_etal:08a} took the solar wind velocity measured by Ulysses at $\sim 5$~AU from the Sun, when it was at the same heliolatitude as Voyager~2 at $\sim 80$~AU, and using a 1D MHD model of the solar wind propagation, they calculated an expected solar wind velocity at Voyager assuming there are no pickup ions. The difference between the modeled and actually measured values, equal to 67~km/s, was adopted as the slowdown of the solar wind due to the ``friction'' of the solar wind against the inflowing neutral interstellar gas. Subsequently, they repeated the calculations of the solar wind speed at Voyager, but this time assuming three values of neutral interstellar hydrogen density at the termination shock: 0.07, 0.9, and 0.11~cm$^{-3}$. Having compared the measured slowdown  value of 67 km/s with modeling results, they arrived at a value of $n_{H,TS} = 0.09 \pm 0.01$~cm$^{-3}$. With the observational uncertainty including potential variations in parameters, such as the ionization rate, and the modeling approach, the overall uncertainty for this derivation can be placed at $\pm 0.02$~cm$^{-3}$. 

A third method was applied by \citet{pryor_etal:08a}, who used UV backscattering observations in an approach that does not require any absolute calibration of the observed UV intensities. They used data from Cassini located at 10 AU, and Voyager at $\sim 90$~AU from the Sun, obtained during the same time interval 2003 -- 2004. Both spacecraft observed about monthly-periodic variations in the backscattering signal. These were attributed to the scattering of the Lyman-alpha radiation from active regions, which rotate with the Sun and thus illuminate the surrounding interstellar gas like a ``lighthouse''. The amplitudes of the variations observed at Voyager were systematically lower than at Cassini, which was interpreted as attenuation by a factor of 0.2, resulting from multiple scattering of the Lyman-alpha photons in the ambient neutral interstellar gas in the inner heliosphere. Using the Moscow MC model of the heliosphere and the radiation transport code by \citet{gangopadhyay_etal:06}, they simulated the attenuation for two sets of interstellar parameters: $(n_p = 0.06 \rm{~cm}^{-3}, n_{\mathrm H} = 0.18 \rm{~cm}^{-3})$ and $(n_p = 0.05 \rm{~cm}^{-3}, n_{\mathrm H} = 0.15 \rm{~cm}^{-3})$, which brought the interstellar gas density at the termination shock equal to 0.085 and 0.095~cm$^{-3}$, respectively. While it was impossible to exactly reproduce all observed variations, they concluded that the attenuation is consistent with a density at the termination shock of $\sim 0.09$~cm${-3}$. However, the value derived is model-dependent. Therefore, an overall uncertainty of about 25\% has been placed on the derived value of $n_{\mathrm{H,TS}} = 0.09 \pm 0.024$~cm$^{-3}$ \citep{pryor_etal:08a}. 

Although aimed primarily at a determination of the ionization rate in the inner heliosphere, another recent study of heliospheric backscatter glow by \citet{quemerais_etal:06b}, which is based on multi-year photometric all-sky observations with SOHO SWAN, has also returned a value of 0.09~cm$^{-3}$ for the H density at the termination shock.

Hence, all three results obtained using independent methods and independent data sets are consistent with each other within their uncertainty bands, and with the result of the PUI analysis by \citet{bzowski_etal:08a}. Computing a weighted average from all these values as done for He by \citet{mobius_etal:04a}, the density of neutral interstellar hydrogen at the termination shock is found equal to $0.089 \pm 0.022$~cm$^{-3}$. 

As an alternative attempt, Gloeckler et al. (2008, this volume) combined the CHISM density of neutral He, based on the combined He parameter determination with a density $n_{\mathrm{He}} = 0.015 \pm 0.0015\pm 0.002$~cm$^{-3}$ \citep{mobius_etal:04a}, with the abundance ratios obtained with Voyager from the suprathermal tails of interstellar pickup ion distributions and from anomalous cosmic rays in the heliosheath. They present a noticeably lower H density at the termination shock of $0.055 \pm 0.021$~cm$^{-3}$, but with a rather large uncertainty due to a number empirical modeling steps involved in the method.

\subsection{Extrapolating the termination shock density to the CHISM}
\label{sec:extrapo}

The bulk velocity and temperature of the gas in the CHISM are now well established from interstellar helium observations \citep{witte:04, gloeckler_etal:04b, vallerga_etal:04a, lallement_etal:04b}, and compiled by \citet{mobius_etal:04a}, but the density, ionization, and magnetic field still require further studies. While the determinations of the density of interstellar H at the termination shock based on various observations, discussed earlier in this section, depend relatively weakly on the modeling of the heliosphere as a whole, extrapolating this density to the CHISM is certainly model-dependent and requires a detailed model of the interaction of the solar wind with the local interstellar medium. Since the mean free path of the interstellar H atoms is comparable with the size of the heliosphere, kinetic models are required. In fact, due to modifications in the interaction region the flow of H atoms at the termination shock varies with the angle off the upwind direction. This dependence should be taken into account in the analyses. In this paper we assumed that the hydrogen interstellar at the termination shock gas consists of two populations (primary and secondary) which flow homogeneously at the TS. It was verified \citep[see][]{bzowski_etal:08a} that such an assumption does not introduce an error greater than 1\% at the location of Ulysses observations discussed in this paper. 

Under these assumptions we can estimate the H atom number density in CHISM by introducing the filtration factor \citep[e.g.][]{izmodenov:07} as: $F = n_{\mathrm{H,TS}}/n_{\mathrm{H,CHISM}}$, where $n_{\mathrm{H,TS}}$ is equal to the density of neutral interstellar hydrogen at the nose of the termination shock and $n_{\mathrm{H,CHISM}}$ to the H density in the unperturbed interstellar gas. Then, knowing $n_{\mathrm{H,TS}}$ upwind and the filtration factor from the models, we can obtain an estimate for $n_{\mathrm{H,CHISM}}$. The filtration factor should be calculated in the framework of a solar wind/CHISM interaction model.

\citet{izmodenov_etal:04a} present an axisymmetric stationary kinetic-gasdynamic model of the solar wind/CHISM interaction. The model differs from the original \citet{baranov_malama:93} by taking into account the solar wind alpha particles and the ionized interstellar helium component. The solar wind parameters in this parametric study were similar to those used by \citet{bzowski_etal:08a}. Table 1 in \citet{izmodenov_etal:04a} presents the filtration factors obtained for different interstellar proton and H atom number densities. It is shown that the filtration factor varies from 0.51 to 0.58 for a wide range of densities. Applying this filtration to the density at the nose of the termination shock obtained by \citet{bzowski_etal:08a} we obtain $n_{\mathrm{H, CHISM}} = 0.16 \pm 0.04$~cm$^{-3}$. However, since the error bars for $n_{\mathrm{H, TS}}$ are large, a wider range of ($n_{\mathrm{H, CHISM}}$, $n_{\mathrm{p, CHISM}}$) is needed than available from \citet{izmodenov_etal:04a}. 

In addition, filtration factors also depend on the interstellar magnetic field, which was not taken into account by \citet{izmodenov_etal:04a}, and on variations in the solar wind ram pressure (\citet{mccomas_etal:08a}; for model, see, e.g., \citet{izmodenov_etal:05a}, \citet{izmodenov_etal:08a}) Also, the dependence of the solar wind on heliolatitude may be important. Including these effects may extend the range of the filtration factor beyond the values mentioned above. To establish the interstellar parameters based on $n_{\mathrm{H, TS}}$, additional constraints, such as the location of the TS in Voyager~1 and 2 directions (see, e.g., Izmodenov, this volume), Lyman-alpha absorption (Wood et al, this volume), and solar backscattered Lyman-alpha radiation observations (Quemerais et al., this volume), should be added.

It is worthwhile to mention that the interaction of interstellar gas with the solar wind is also described  with hydrodynamic multi-fluid models \citep[e.g.][]{mueller_etal:06, scherer_fahr:03a, florinski_etal:05a, florinski_etal:03a}. \citet{alexashov_izmodenov:05a} compared results of their kinetic model with different multi-fluid models. Results relevant to present studies are listed in Table~2 of this paper and show that the parameters of the primary and secondary populations of interstellar atoms obtained from hydrodynamic models differ significantly from the results of the kinetic models.  

\citet{mueller_etal:08a} compared the results from five different models of the heliosphere evaluated with identical boundary conditions. Two of these models are fully kinetic in their treatment of neutrals \citep{baranov_malama:93,heerikhuisen_etal:06a}, and three are hydrodynamic multi-fluid (\citet{florinski_etal:03a,florinski_etal:05a};  \citet{mueller_etal:06}, extended from \citet{pauls_etal:95}, and \citet{scherer_fahr:03a}, extended from \citet{fahr_etal:00}). The parameter set used by \citet{mueller_etal:08a} was close to one of the sets of interstellar parameters used in the simulations by \citet{bzowski_etal:08a} with the important exception that the solar wind speed was essentially lower. The filtration factors obtained in this comparison varied from 0.52 (for the two-fluid hydrogen model) to 0.69-0.74 (for the kinetic models). 

\citet{mueller_etal:08a} did not present the parameters of the primary and secondary populations at the termination shock separately and offered only the mean values of the density, bulk velocity, and temperature. While all models reproduced the mean bulk velocity at the termination shock  within a narrow range from 19.2 to 23.4~km/s, differences in the temperature were larger, i.e. from 12~000 to 30~900 K.

Since the goal of the comparison made by \citet{mueller_etal:08a} was checking the results of different calculating schemes, all models had to include the same physical effects, which required switching off some of their advanced features (different in different models). As a result, none of them was run in its full mode. Nevertheless, both of the comparison studies (\citet{alexashov_izmodenov:05a} and \citet{mueller_etal:08a}) seem to suggest that an uncritical use of hydrodynamic multi-fluid models for quantitative analysis of  heliospheric observations may be misleading.

From the discussion above it follows that without performing a dedicated modeling, the H density in the CHISM can be assessed from the density at the termination shock by dividing this value by the filtration ratio being in the (0.51, 0.59) range, which brings the density of neutral H in the CHISM equal to $0.16 \pm 0.04$~cm$^{-3}$. 

\section{Conclusions}
\label{sec:conclu}
The density of neutral interstellar gas at the termination shock seems to be well constrained by observations performed using different techniques and appropriate modeling. The datasets used include Ulysses observations of H pickup ions, Voyager and Ulysses observations of solar wind slowdown, Cassini and Voyager observations of modulation in the heliospheric Lyman-alpha glow, and multi-year all-sky photometric observations of the glow by SWAN/SOHO. The densities obtained from analysis of these observations group in the range of 0.08 -- 0.09~cm$^{-3}$ and the uncertainty of this determination is about 20\%. Extrapolation of this value to the CHISM based on analysis of heliospheric filtration factors from the literature brings the density of neutral interstellar hydrogen equal to $0.16 \pm 0.04$~cm$^{-3}$.

\begin{acknowledgements}
The authors gratefully acknowledge the hospitality and excellent ambiance within the International Space Science Institute (ISSI) in Bern, Switzerland, where the workshop "From the Outer Heliosphere to the Local Bubble: Comparison of New Observations with Theory" was held. This research was performed within the framework of an International Space Science Institute (Bern, Switzerland) Working Group Neutral Interstellar Hydrogen. The authors are grateful for helpful discussions during the manuscript preparations with Martin Lee and Stan Grzedzielski. Support for this study from NASA Grant NNG06GF55G and Grant NAG 5-12929 through a subcontract from the California Institute of Technology is gratefully acknowledged. V.I. was supported part by RFBR grants 07-02-01101-a, 07-01-00291-a and Dynastia Foundation.
\end{acknowledgements}

\newcommand{\bibfont}{\footnotesize}
\bibliographystyle{SSRv}
\bibliography{iplbib} 

\end{document}